\documentclass{llncs}
\usepackage{verbatim}
\usepackage{amssymb}
\usepackage{rotating}
\usepackage{algorithm,algorithmic}
\usepackage{tabularx}
\usepackage{subfigure}

\begin{document}
\title{Computing Strong Game-Theoretic \\Strategies in Jotto\thanks{This material is based upon work supported by the National Science
Foundation under grants IIS-0964579, IIS-0905390, and CCF-1101668.}}
\author{Sam Ganzfried}
\institute{Carnegie Mellon University \\ Computer Science Department \\
\email{sganzfri@cs.cmu.edu}}

\date{}

\maketitle

\begin{abstract}
We develop a new approach that computes approximate \linebreak equilibrium strategies in Jotto, a popular word game.  Jotto is an extremely large two-player game of imperfect information; its game tree has many orders of magnitude more states than games previously studied, including no-limit Texas hold 'em. To address the fact that the game is so large, we propose a novel strategy representation called oracular form, in which we do not explicitly represent a strategy, but rather appeal to an oracle that quickly outputs a sample move from the strategy's distribution.  Our overall approach is based on an extension of the fictitious play algorithm to this oracular setting. We demonstrate the superiority of our computed strategies over the strategies computed by a benchmark algorithm, both in terms of head-to-head and worst-case performance. 
\end{abstract}

\section{Introduction}
\label{Intro}
Developing strong strategies for agents in large games is an important problem in artificial intelligence.  In particular, much work has been devoted in recent years to developing algorithms for computing game-theoretic solution concepts, specifically the Nash equilibrium.  In two-player zero-sum games, Nash equilibrium strategies have a strong theoretical justification as they also correspond to minimax strategies; by following an equilibrium strategy, a player can guarantee at least the value of the game in expectation, regardless of the strategy followed by his opponent.  Currently, the best algorithms for computing a Nash equilibrium in two-player zero-sum extensive-form games (with perfect recall) are able to solve games with $10^{12}$ states in their game tree~\cite{Zinkevich07:Regret}.    

Unfortunately, many interesting games actually have significantly more than $10^{12}$ states. Texas hold 'em poker is a prime example of such a game that has received significant attention in the AI literature in recent years; the game tree of two-player limit Texas hold 'em has about $10^{18}$ states, while that of two-player no-limit Texas hold 'em has about $10^{71}$ states. The standard approach of dealing with this is to apply an \emph{abstraction} algorithm, which constructs a smaller game that is similar to the original game; then the smaller game is solved, and its solution is mapped to a strategy profile in the original game~\cite{Billings03:Approximating}. Many abstraction algorithms work by coarsening the moves of chance, collapsing several information sets of the original game into single information sets of the abstracted game (called \emph{buckets}).

In this paper we study a game with many orders of magnitude more states than even two-player no-limit Texas hold 'em.  Jotto, a popular word game, contains approximately $10^{853}$ states in its game tree.  Unfortunately, Jotto does not seem particularly amenable to abstraction in the same way that poker is; we discuss reasons for this in Section~\ref{sect:Jotto}.  Furthermore, even if we could apply an abstraction algorithm to Jotto, we would need to group $10^{841}$ game states into a single bucket on average, which would almost certainly lose a significant amount of information from the original game.  Thus, the abstraction paradigm that has been successful on poker does not seem promising to games like Jotto; an entirely new approach is needed.  

We provide such an approach.  To deal with the fact that we cannot even represent a strategy for one of the players, we provide a novel strategy representation which we call \emph{oracular form}.  Rather than viewing a strategy as an explicit object that must be represented and stored, we instead represent it implicitly through an oracle; we can think of the oracle as an efficient algorithm.  Each time we want to make a play from the strategy, we query the oracle, which quickly outputs a sample play from the strategy's distribution.  Thus, instead of representing the entire strategy in advance, we obtain it on an as-needed basis via real-time computation.         

Our main algorithm for computing an approximate equilibrium in Jotto is an extension of the fictitious play algorithm~\cite{Brown51:Iterative} to our oracular setting.  The algorithm outputs a full strategy for one player, and for the other player outputs data such that if another algorithm is run on it, a sample of the strategy's play is obtained.  Thus, we can play this strategy, even though it is never explicitly represented.  We use our algorithm to compute approximate equilibrium strategies on 2, 3, 4, and 5-letter variants of Jotto. We demonstrate the superiority of our computed strategies over the strategies computed by a benchmark algorithm, both in terms of head-to-head and worst-case performance. 

\section{Jotto}
\label{sect:Jotto}
\begin{figure}[!ht]
\begin{center}
\includegraphics[scale = 0.5] {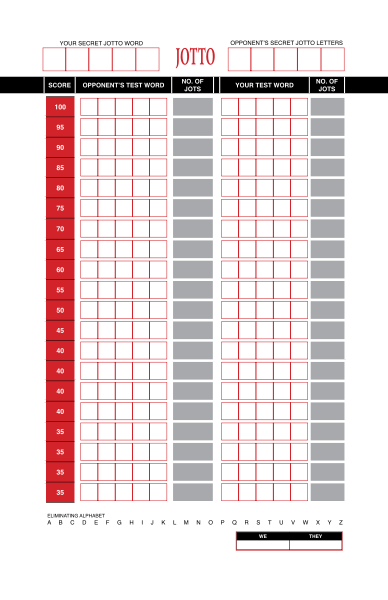}
\vspace{-0.3in}
\caption{Official Jotto sheet. Players record guesses and answers, cross off alphabet letters as they become inconsistent, and record other notes and deductions.}
\label{fig:Jotto-sheet}
\end{center}
\vspace{-0.3in}
\end{figure}

Jotto is a popular two-player word game.  While there are many different variations of the game, we will describe the rules of one common variant.  Each player picks a secret five-letter word, and the object of the game is to correctly guess the other player's word first.  Players take turns guessing the other player's secret word and replying with the number of common letters between the guessed word and the secret word (the positions do not matter).  For example, if the secret word is GIANT and a player guesses PECAN, the other player will give a reply of 2 (for the A and the N, even though they are in the wrong positions).  Players often cross out letters that are eliminated and record other logical deductions on a sheet of paper.  An official Jotto sheet is shown in Figure~\ref{fig:Jotto-sheet}.

Instead of having both players simultaneously guessing the other player's word, we could instead just have one player pick a secret word while the other player guesses it.  Let us refer to these players as the \emph{hider} and the \emph{guesser} respectively.  If the guesser correctly guesses the word on his $k$'th try, then the hider gets payoff $k$ while the guesser gets payoff $-k.$  This is the variation that we consider in the remainder of the paper.

There are a few limits on the words that the players can select.  All words must be chosen from a pre-arranged dictionary.  No proper nouns are allowed, and the words must consist of all different letters (some variations do not impose this restriction).  Furthermore, we do not allow players to select a word of which other permutations (aka anagrams) are also valid words (e.g., STARE and RATES)\footnote{If this restriction were not imposed, then players might need to guess all possible permutations of a word even once all the letters are known.}. 

The official dictionary we will be using has 2833 valid 5-letter words.  A na\"{\i}ve attempt at determining the size of the game tree is the following. First the hider selects his word, putting the game into one of 2833 states.  At each state, the guesser must choose one of 2833 words; the hider gives him an answer from 0-5, and the guesser must now choose one of 2832 words, and so on.  The total number of game states will be approximately $2833 \cdot 2833!.$
  
It turns out that we can represent the game much more concisely if we take advantage of the fact that many paths of play lead to the guesser knowing the exact same information.  For example, if the player guesses GIANT and gets reply of 2 followed by guessing PECAN with a reply of 3, he knows the exact same information as if he had guessed PECAN with a reply of 3 followed by GIANT with a reply of 2; both sequences should lead to the same game state.  More generally, two sequences of guesses and answers lead to identical knowledge bases if and only if the set of words consistent with both sequences is identical.  Thus, there is a one-to-one correspondence between a knowledge base of the guesser and a subset of the set of words in the dictionary corresponding to the set of words that are consistent with the guesses so far.  Since there are $2^{2833}$ total subsets of words in the dictionary, the total number of game states where the guesser would need to make a decision is $2^{2833} \approx 10^{853}.$  Since the best equilibrium-finding algorithms can only solve games with up to $10^{12}$ nodes, we have no hope of solving Jotto by simply applying an existing algorithm.

Furthermore, Jotto does not seem amenable to the same abstraction para\-digm that has been successful on poker. In poker, abstraction works by grouping several states into the same \emph{bucket} and forcing all states in the same bucket to follow the same strategy.  In our compact representation of Jotto, the states correspond to subsets of the dictionary; abstraction would mean that several subsets are grouped together into single buckets.  However, the action taken at each bucket will be a single word (i.e., the next guess).  If many subsets are grouped together into the same bucket, then there will clearly be some words that have already been guessed in some states in the bucket, while not in other states.  This will lead to certain actions being ill-defined, as well as possible infinite loops in the structure of the abstracted game tree.  This will be exacerbated by the fact that many states will need to be grouped into the same bucket; on average, each bucket will contain $10^{841}$ game states. 

In short, there are significant challenges that must be overcome to apply any sort of abstraction to Jotto; this may not even be feasible to do at all.  Instead, we propose an entirely new approach.

\section{A natural approach}
\label{sect:benchmark}
A natural strategy for the hider would be to select each word uniformly at random, and one for the guesser would be to always guess the word that will eliminate the most words that are still consistent with the guesses and answers so far (in expectation against the uniform hider strategy). We refer to this strategy for the hider as HiderUniform, and to the strategy for the guesser as GuesserGBRUniform (for ``Greedy Best Response'').  GuesserGBRUniform is essentially 1-ply minimax search; further details and pseudocode are given in Section~\ref{sect:GBR}. We suspect that current Jotto programs follow algorithms similar to HiderUniform and GuesserGBRUniform, though we are not aware of any publicly available descriptions of existing algorithms. We will use these algorithms as a benchmark to measure the performance of our new approach.

While HiderUniform seems like a pretty strong (i.e., low-exploitabil\-ity) strategy for the hider, it turns out that GuesserGBRUniform is actually highly exploitable.  For example, in the five-letter variant using our rules and dictionary, GuesserGBRUniform will always select `doyen' as its first guess.  Clearly no intelligent hider would select doyen as his secret word against such an opponent.  Furthermore, the hider can always guarantee that GuesserGBRUniform will require 9 guesses by selecting a word such as `amped' (note that 9 is the maximal number of guesses that GuesserGBRUniform will take to guess any word in the 5-letter variant).    

Searching additional levels down the tree will probably not help much with the worst-case exploitability of the guesser's strategy.  The main problem is that GuesserGBRUniform plays a deterministic strategy, and a worst-case opponent could exploit it by always selecting the word that requires the most guesses.  We would like to compute a less exploitable strategy for the guesser, which will involve some amount of randomization.  Our overall goal is to compute strategies for both players with worst-case exploitabilities as low as possible (i.e., we would like to compute an approximate Nash equilibrium, viewing Jotto as a game of imperfect information). 

\section{Game theory background}
\label{sect:gt}
In this section, we review relevant definitions and prior results from game theory and game solving.

\subsection{Strategic-form games}
The most basic game representation, and the standard representation for simul\-taneous-move games, is the \emph{strategic form}.
A \emph{strategic-form game} (aka matrix game) consists of a finite set of players $N,$ a space of \emph{pure strategies} $S_i$ for each player, and a utility function $u_i: \times_i S_i \rightarrow \mathbb{R}$ for each player.  Here $\times_i S_i$ denotes the space of \emph{strategy profiles} --- vectors of pure strategies, one for each player.

The set of \emph{mixed strategies} of player $i$ is the space of probability distributions over his pure strategy space $S_i$.  We will denote this space by $\Sigma_i.$  If the sum of the payoffs of all players equals zero at every strategy profile, then the game is called \emph{zero sum}.  In this paper, we will be primarily concerned with two-player zero-sum games. If the players are following strategy profile $\sigma,$ we let $\sigma_{-i}$ denote the strategy taken by the opponent.

\subsection{Extensive-form games}
An \emph{extensive-form} game is a general model of multiagent decision-making with potentially sequential and simultaneous actions and imperfect information.  As with perfect-information games, extensive-form games consist primarily of a game tree; each non-terminal node has an associated player (possibly \emph{chance}) that makes the decision at that node, and each terminal node has associated utilities for the players.  Additionally, game states are partitioned into \emph{information sets}, where the player whose turn it is to move cannot distinguish among the states in the same information set.  Therefore, in any given information set, the player whose turn it is to move must choose actions with the same distribution at each state contained in the information set.  If no player forgets information that he previously knew, we say that the game has \emph{perfect recall}. 

\subsection{Mixed vs. behavioral strategies}
There are multiple ways of representing strategies in extensive-form games.  Define a \emph{pure strategy} to be a vector that specifies one action at each information set.  Clearly there will be an exponentially number of pure strategies in the size of the game tree; if the tree has $I_i$ information sets for player $i$ and $A_i$ possible actions at each information set, then player $i$ has $A_i ^ {I_i}$ possible pure strategies.  Define a \emph{mixed strategy} to be a probability distribution over the space of pure strategies.  We can represent a mixed strategy as a vector with $A_i ^ {I_i}$ components.  

Alternatively, we could play a strategy that randomizes independently at each information set; we refer to such a strategy as a \emph{behavioral strategy}.  Since a behavioral strategy must specify a probability for playing each of $A_i$ actions at each information set, we can represent it as a vector with only $A_i \cdot I_i$ components.  Thus, behavioral strategies can be represented exponentially more compactly than mixed strategies. 

Fortunately, it turns out that this gain in representation size does not come at the loss of expressiveness; any mixed strategy can also be represented as an equivalent behavioral strategy (and vice versa). 
Thus, current computational approaches to extensive-form games operate on behavioral strategies and avoid the unnecessary exponential blowup associated with using mixed strategies. 

\subsection{Nash equilibria}
\label{sect:ne}
Player $i$'s \emph{best response} to $\sigma_{-i}$ is any strategy in $\mbox{arg}\max _{\sigma'_i \in \Sigma_i} u_i(\sigma'_i, \sigma_{-i}).$  A \emph{Nash equilibrium} is a strategy profile $\sigma$ such that $\sigma_i$ is a best response to $\sigma_{-i}$ for all $i.$  An $\epsilon$-\emph{equilibrium} is a strategy profile in which each player achieves a payoff of within $\epsilon$ of his best response.  

In two player zero-sum games, we have the following result which is known as the \emph{minimax theorem}:
$$v^* = \max_{\sigma_1 \in \Sigma_1} \min_{\sigma_2 \in \Sigma_2} u_1(\sigma_1,\sigma_2) = \min_{\sigma_2 \in \Sigma_2} \max_{\sigma_1 \in \Sigma_1} u_1(\sigma_1,\sigma_2).$$
We refer to $v^*$ as the \emph{value} of the game to player 1.  
Any equilibrium strategy for a player guarantees an expected payoff of at least the value of the game to that player.  

All finite games have at least one Nash equilibrium.  Currently, the best algorithms for computing a Nash equilibrium in two-player zero-sum extensive-form games with perfect recall are able to solve games with $10^{12}$ states in their game tree~\cite{Zinkevich07:Regret}.    

\section{Smoothed fictitious play}
\label{sect:sfp}
In this section we will review the fictitious play (FP) algorithm~\cite{Brown51:Iterative}. Despite its conceptual simplicity, FP has recently been used to compute equilibria in many classes of games in the artificial intelligence literature (e.g.,~\cite{Ganzfried08:Computing,Rabinovich09:Generalised}).     
The basic FP algorithm works as follows. At each iteration, each player plays a best response to the average strategy of his opponent so far (we assume the game has two players). Formally, in smoothed fictitious play each player $i$ applies the following update rule at each time step $t$: 
$$s_{i,t} = \left( 1 - \frac{1}{t+1} \right) s_{i,t-1} + \frac{1}{t+1} s^{BR}_{i,t},$$
where $s^{BR}_{i,t}$ is a best response of player $i$ to the strategy $s_{-i,t-1}$ of his opponent at time $t-1.$  We allow strategies to be initialized arbitrarily at $t = 0.$

FP is guaranteed to converge to a Nash equilibrium in two-player zero-sum games; however, very little is known about how many iterations are needed to obtain convergence. Recent work shows that FP may require exponentially many iterations in the worst case~\cite{Brandt10:On}; however, it may perform far better in practice on specific games.  In addition, the performance of FP is not monotonic; for example, it is possible that the strategy profile after 200 iterations is actually significantly closer to equilibrium than the profile after 300 iterations. 
So simply running FP for some number of iterations and using the final strategy profile is not necessarily the best approach.

We instead use the following improved algorithm.  For each iteration we compute the amount each player could gain by deviating to a best response; denote it by $\epsilon_{i,t}$.  Let $\epsilon_{t} = \max_i \epsilon_{i,t},$ and let $\epsilon_{t^*} = \min_{0 \leq t \leq T} \epsilon_{t}.$  After running FP for $T$ iterations, rather than output $s_{i,T},$ we will instead output $s_{i,t^*}$ --- the $\epsilon$-equilibrium for smallest $\epsilon$ out of all all the iterations of FP so far.

\section{Oracular strategies}
\label{sect:orac_form}
Consider the following scenario.  Suppose one is playing an extensive-form game $G$ with $2^{100}$ information sets and two actions per information set, and that he wishes to play an extremely simple strategy: always choose the first action at each information set (suppose actions are labeled as Action 1 and Action 2).  To represent this pure strategy, technically we must list out a vector of size $2^{100}$ (with each entry being a 1 for this particular strategy).  On the other hand, it is trivial to write an algorithm that takes as input the index of an information set and outputs the action taken by this strategy (i.e., output 1 on all inputs).  Even though there are a large number of information sets, we only require 100 bits to represent the index of each one; thus, it is possible to play this simple strategy without ever explicitly representing it.    

More generally, let $O_i$ be an efficient deterministic algorithm that takes as input the index of an information set $I$ and outputs an action from $A_I,$ the set of actions available at $I.$  We refer to $O_i$ as a \emph{pure oracular strategy} for player $i$.  It is easy to see that every pure oracular strategy is strategically equivalent to a pure strategy $s_i$ of the game; at information set I, $s_i$ simply plays whatever action $O_i$ outputs on input $I.$  

We define oracular versions of randomized strategies analogously to the extensive-form case.  Let $\{O_i\}$ be a collection of pure oracular strategies; then any probability distribution over elements of $\{O_i\}$ is a \emph{mixed oracular strategy}. If we let $O_i$ be a randomized algorithm that outputs a probability distribution over actions at each information set, then call $O_i$ a \emph{behavioral oracular strategy}.  As was the case with pure strategies, each oracular strategy corresponds to a single extensive-form strategy of the same type.   

In the next two sections, we will see how the oracular strategy representation can be useful in practice when computing approximate equilibrium strategies in Jotto.  In particular, a strategy for the guesser is so large that we cannot represent it explicitly; however, we can encode it concisely as an oracular strategy which we efficiently query repeatedly throughout the algorithm.       

\section{Computing best responses in Jotto}
\label{sect:JottoBR}
In order to apply smoothed fictitious play to Jotto, we must figure out how to compute a best response for each player.  This is challenging for several reasons.  First, the guesser's strategy space is so large that we cannot compute a full best response; we must settle for computing an approximate best response, which we call the guesser's \emph{greedy best response}.  In addition, we represent the guesser's strategy in oracular form; so the hider can cannot operate on it explicitly, and can only query it at certain game states.  It turns out that we can actually compute an exact best response for the hider despite this limitation. 

\subsection{Computing the guesser's greedy best response}
\label{sect:GBR}
Suppose we are given a strategy for the hider, and wish to compute a counterstrategy for the guesser.  Let $h$ denote the strategy of the hider, where $h_i$ denotes the probability that the hider chooses $w_i$ --- the $i$'th word in the dictionary. Let $D$ denote the number of words in the dictionary, and let $S$ be a bit-vector of size $D,$ where $S_i = 1$ means that $w_i$ is still consistent with the guesses so far.  So $S$ encodes the current knowledge base of the guesser and represents the state of the game.    

A reasonable heuristic to use for the guesser would be the following.  For each word $w_i$ in the dictionary, compute the number of words that we will eliminate in expectation (over $h$) if we guess $w_i.$  Then guess the word that expects to eliminate the most words.  We refer to this algorithm as GuesserGBR (for ``Greedy Best Response''); pseudocode is given in Algorithm~\ref{alg:guesser-gbr}.

GuesserGBR relies on a number of subroutines.  ExpNumElims, given in Algorithm~\ref{alg:exp_num_elim}, gives the expected number of words eliminated if $w_i$ is guessed.  AnswerProbs, given in Algorithm~\ref{alg:ans_probs}, gives a vector of the expected probability of receiving each answer from the hider when $w$ is guessed.  NumElims, given in Algorithm~\ref{alg:num_elim}, gives the number of words that can be eliminated when $w_i$ is guessed and an answer of $j$ is given.  Finally, NumCommLetts, given in Algorithm~\ref{alg:num_comm_lett}, gives the number of common letters between two words.  In the pseudocode, $L$ denotes the number of letters allowed per word.  For efficiency, we precompute a table of size $D^2$ storing all the numbers of common letters between pairs of words.  The overall running time of GuesserGBR is $O(D^2 L).$  

It is worth noting that the greedy best response is not an actual best response; it is akin to searching one level down the game tree and then using the evaluation function ``expected number of words eliminated'' to determine the next move.  This is essentially 1-ply minimax search.  While we would like to compute an exact best response by searching the entire game tree, this is not feasible since the tree has $10^{853}$ nodes.  As with computer chess programs, we will need to settle for searching down the tree as far as we can, then applying a reasonable evaluation function.

\begin{figure*}[ttt!]
\begin{minipage}[t]{2.3in}
\begin{algorithm}[H]
\caption{\small GuesserGBR($h$, $S$)}
\scriptsize
\begin{algorithmic}
\FOR {$i = 1$ to $D$}
\STATE $n_i \gets ExpNumElims(w_i, h, S)$
\ENDFOR 
\RETURN $w_i$ with maximal value of $n_i$
\end{algorithmic}
\label{alg:guesser-gbr}
\end{algorithm}
\vspace{-0.6in}
\begin{algorithm}[H]
\caption{\small ExpNumElims($w_i$, $h$, $S$)}
\scriptsize
\begin{algorithmic}
\STATE $A \gets AnswerProbs(w_i, h, S)$
\STATE $n \gets 0$
\FOR {$j = 1$ to $L$}
\STATE $n \gets n + A_j \cdot NumElims(w_i, S, j)$
\ENDFOR
\RETURN $n$
\end{algorithmic}
\label{alg:exp_num_elim}
\end{algorithm}
\vspace{-0.6in}
\begin{algorithm}[H]
\caption{\small NumElims($w_i$, $S$, $j$)}
\scriptsize
\begin{algorithmic}
\STATE counter $\gets$ 0
\FOR {$k = 1$ to $D$}
\IF {$S_k = 1$ and \\ $NumCommLetts(w_i,w_k) \neq j$}
\STATE counter $\gets$ counter + 1
\ENDIF
\ENDFOR
\RETURN counter
\end{algorithmic}
\label{alg:num_elim}
\end{algorithm}
\end{minipage}
\hfill
\begin{minipage}[t]{2.3in}
\begin{algorithm}[H]
\caption{\small AnswerProbs($w$, $h$, $S$)}
\scriptsize
\begin{algorithmic}
\FOR {$i = 1$ to $L$}
\STATE $A_i \gets 0$
\ENDFOR
\FOR {$i = 1$ to $D$}
\IF {$S_i = 1$}
\STATE $k \gets NumCommLetts(w, w_i)$
\STATE $A_k \gets A_k + h_i$
\ENDIF
\ENDFOR
\STATE Normalize $A$ so its elements sum to 1.
\RETURN $A$
\end{algorithmic}
\label{alg:ans_probs}
\end{algorithm}
\vspace{-0.6in}
\begin{algorithm}[H]
\footnotesize
\caption{\small NumCommLetts($w_i$, $w_j$)}
\scriptsize
\begin{algorithmic}
\STATE counter $\gets$ 0
\FOR {$m = 1$ to $L$}
\STATE $c_1 \gets$ $m$th character of $w_i$
\FOR {$n = 1$ to $L$}
\STATE $c_2 \gets$ $n$th character of $w_j$ 
\IF {$c_1 = c_2$}
\STATE $counter \gets counter + 1$
\ENDIF
\ENDFOR
\ENDFOR
\RETURN counter
\end{algorithmic}
\label{alg:num_comm_lett}
\end{algorithm}
\end{minipage}
\hfill
\end{figure*}

\subsection{Computing the hider's best response}
\label{sect:hider-br}
In order to compute the hider's best response (in the context of fictitious play), we will find it useful to introduce two data structures. Let IterNumGuesses (ING) and AvgNumGuesses (ANG) be two arrays of size $D$.  The $i$'th component of ING will denote the number of guesses needed for the guesser's greedy best response to guess $w_i$ at the current iteration of the algorithm. The $i$'th component of ANG will be the average over all iterations (of fictitious play) of the number of guesses needed for the guesser's greedy best response to guess $w_i.$ We update ANG by applying 
$$ANG[i] = \left( 1 - \frac{1}{t+1} \right) ANG[i] + \frac{1}{t+1} ING[i].$$

We update ING at each time step by applying 
$$ING = CompNumGuesses(s_{h,t}),$$ 
where $s_{h,t}$ is the hider's strategy at iteration $t$ of fictitious play, and pseudocode for CompNum\-Guesses is given below in Algorithm~\ref{alg:compute-num-guesses}.  CompNumGuesses computes the number of guesses needed for the guesser's greedy best response to $s_{h,t}$ to correctly guess each word.  It accomplishes this by repeatedly querying GuesserGBR at various game states.  The subroutine UpdateState updates the game state in light of the answer received from GuesserGBR.  It is in this way that the hider's best response algorithm selectively queries the guesser's strategy, which is represented implicitly in oracular form.

Finally, once ING and ANG have been updated as described above, we are ready to compute the full best response for the hider. Pseudocode for the algorithm HiderBR is given below in Algorithm~\ref{alg:hider-br}.  HiderBR takes ANG as input, and determines which word(s) required the most guesses on average. If there is a unique word requiring the maximal number of guesses, then that word is selected with probability 1. If there are multiple words requiring the maximal number of guesses, then these are each selected with equal probability.  Note that selecting any distribution over these words would constitute a best response; we just choose one such distribution.  The asymptotic running time of CompNumGuesses is $O(D^4 L),$ while that of HiderBR is $O(D).$        
     
Note that, unlike GuesserGBR of Section~\ref{sect:GBR} which is an approximate best response using 1-ply minimax search, HiderBR is a full best response. We are able to compute a full best response for the hider because his strategy space is much smaller than that of the guesser; the hider has only $D$ possible pure strategies --- 2833 in the case of 5-letter Jotto.      

\begin{figure*}[ttt!]
\begin{minipage}[t]{2.3in}
\begin{algorithm}[H]
\caption{\small CompNumGuesses($s_h$)}
\scriptsize
\begin{algorithmic}
\FOR {$i = 1$ to $D$}
\STATE $S \gets$ vector of size $D$ of all ones.
\STATE $numguesses \gets 0$
\WHILE {TRUE}
\STATE $numguesses \gets numguesses + 1$
\STATE $nextguess \gets GuesserGBR(s_h,S)$
\STATE $answer \gets$ \\ $NumCommLetts(w_i, nextguess)$
\IF {$answer = L$}
\STATE BREAK
\ENDIF
\STATE $S \gets$ \\ $UpdateState(S, nextguess, answer)$
\ENDWHILE
\STATE $x_i \gets numguesses$
\ENDFOR
\RETURN $x$
\end{algorithmic}
\label{alg:compute-num-guesses}
\end{algorithm}
\end{minipage}
\hfill
\begin{minipage}[t]{2.3in}
\begin{algorithm}[H]
\caption{\small \newline UpdateState($S$, $nextguess$, $answer$)}
\scriptsize
\begin{algorithmic}
\FOR {$i = 1$ to $D$}
\IF {$S_i = 1$ and NumCommLetts($nextguess$, $w_i$) $\neq$ answer}
\STATE $S_i \gets 0$
\ENDIF
\ENDFOR
\RETURN $S$
\end{algorithmic}
\label{alg:update-state}
\end{algorithm}
\vspace{-0.6in}
\begin{algorithm}[H]
\caption{\small HiderBR(ANG)}
\scriptsize
\begin{algorithmic}
\STATE $x^* \gets \max_i ANG[i].$
\STATE $T^* \gets \{i : ANG[i] = x^*\}$
\FOR {$i = 1$ to $D$}
\IF {$ANG[i] = x^*$}
\STATE $y_i = \frac{1}{|T^*|}$
\ELSE
\STATE $y_i = 0$
\ENDIF
\ENDFOR
\RETURN $y$
\end{algorithmic}
\label{alg:hider-br}
\end{algorithm}
\end{minipage}
\hfill
\end{figure*}

\subsection{Parallelizing the best response calculation}
\label{sect:parallel}
The hider's best response calculation can be sped up dramatically by parallelizing over several cores. In particular, we parallelize the CompNumGuesses subroutine as follows.  For the first processor, we iterate over $i = 1$ to $\left\lfloor \frac{D}{P} \right\rfloor,$ where $P$ is the number of processors, and so on for the other processors.  Thus we can perform $P$ independent computations in parallel. The overall running time of the new algorithm is $O\left(\frac{D^4 L}{P} \right)$.  

\section{Computing an approximate equilibrium in Jotto}
\label{solving-jotto}
We would like to apply smoothed fictitious play to Jotto, using HiderBR for the hider's best response and GuesserGBR for the guesser's best response; however, this is tricky for several reasons.  First, it is not clear how to compute the epsilons and determine the quality of our strategies.  And second, it will be difficult to run the algorithm without explicitly represent the guesser's strategy; furthermore, we cannot output it at the end of the algorithm. 

It turns that using the data structures developed in Section~\ref{sect:hider-br}, we can actually compute the epsilons quite easily.  This is accomplished using the procedures given in Algorithms~\ref{alg:hider-br-payoff}--~\ref{alg:guesser-actual-payoff}.

We are now ready to present our full algorithm for computing an approximate equilibrium in Jotto; pseudocode is given in Algorithm~\ref{alg:solve-jotto}.  Note that we initialize the hider's strategy to choose each word uniformly at random.  In terms of the guesser's strategy, it turns out that all the information needed to obtain it is already encoded in the hider's strategy and that we do not actually need to represent it in the course of algorithm.  

\begin{figure*}[ttt!]
\begin{minipage}[t]{2.3in}
\begin{algorithm}[H]
\caption{\small HiderBRPayoff(ANG)}
\scriptsize
\begin{algorithmic}
\STATE maxnumguesses $\gets 0$
\FOR {$i = 1$ to $D$}
\IF {ANG[i] $>$ maxnumguesses}
\STATE maxnumguesses $\gets$ ANG[i]
\ENDIF
\ENDFOR
\RETURN maxnumguesses
\end{algorithmic}
\label{alg:hider-br-payoff}
\end{algorithm}
\vspace{-0.6in}
\begin{algorithm}[H]
\caption{\small \newline HiderActualPayoff(ANG, $s_h$)}
\scriptsize
\begin{algorithmic}
\STATE result $\gets 0$
\FOR {$i = 1$ to $D$}
\STATE $result \gets result + ANG[i] \cdot s_h[i]$
\ENDFOR
\RETURN result
\end{algorithmic}
\label{alg:hider-actual-payoff}
\end{algorithm}
\vspace{-0.6in}
\begin{algorithm}[H]
\caption{\small \newline GuesserBRPayoff(ING, $s_h$)}
\scriptsize
\begin{algorithmic}
\STATE result $\gets 0$
\FOR {$i = 1$ to $D$}
\STATE $result \gets result + ING[i] \cdot s_h[i]$
\ENDFOR
\RETURN -1 $\cdot$ result
\end{algorithmic}
\label{alg:guesser-br-payoff}
\end{algorithm}
\vspace{-0.6in}
\begin{algorithm}[H]
\caption{\small \newline GuesserActualPayoff(\small ANG, $s_h$)}
\scriptsize
\begin{algorithmic}
\RETURN -1 $\cdot$ HiderActualPayoff(ANG, $s_h$)
\end{algorithmic}
\label{alg:guesser-actual-payoff}
\end{algorithm}
\end{minipage}
\hfill
\begin{minipage}[t]{2.3in}
\begin{algorithm}[H]
\caption{\small SolveJotto($T$)}
\scriptsize
\begin{algorithmic}
\STATE $s_{h,0} \gets (\frac{1}{D},\ldots,\frac{1}{D})$
\STATE Output $s_{h,0}$ to StrategyFile
\STATE $ING \gets ComputeNumGuesses(s_{h,0})$ 
\STATE $ANG \gets ING$
\STATE Compute $\epsilon$'s per Algorithms~\ref{alg:hider-br-payoff}-\ref{alg:guesser-actual-payoff}, $t^* \gets 0$
\FOR {$t = 1$ to $T$}
\STATE $s^{BR}_{h,t} \gets HiderBR(ANG)$
\STATE $s_{h,t} \gets \left( 1 - \frac{1}{t+1} \right) s_{h,t-1} + \frac{1}{t+1} s^{BR}_{h,t}$   
\STATE Output $s_{h,t}$ to StrategyFile
\STATE $ING \gets ComputeNumGuesses(s_{h,t})$
\STATE $ANG \gets \left( 1 - \frac{1}{t+1} \right) ANG + \frac{1}{t+1} ING$
\STATE Compute $\epsilon$'s per Algorithms~\ref{alg:hider-br-payoff}-\ref{alg:guesser-actual-payoff}
\IF {$\epsilon < \epsilon^*$}
\STATE $\epsilon^* \gets \epsilon$, $t^* \gets t$
\ENDIF
\ENDFOR
\RETURN $(s_{h,t^*}, t^*, StrategyFile)$
\end{algorithmic}
\label{alg:solve-jotto}
\end{algorithm}
\vspace{-0.6in}
\begin{algorithm}[H]
\caption{ObtainGuesserStrategy(\small StrategyFile, $t^*$, $S$)}
\scriptsize
\begin{algorithmic}
\STATE $i \gets randint(1,t^*)$
\STATE $s_h \gets$ strategy vector on $i$'th line of StrategyFile
\RETURN GuesserGBR($s_h$, $S$)  
\end{algorithmic}
\label{alg:obtain-guesser-strategy}
\end{algorithm}
\end{minipage}
\hfill
\end{figure*}

To obtain the guesser's final strategy, note that the strategies of the hider are output to a file at each iteration.  It turns out that we can use this file to efficiently generate samples from the guesser's strategy, even though we never explicitly output this strategy.  We present pseudocode in Algorithm~\ref{alg:obtain-guesser-strategy} for generating a sample of the guesser's strategy at state $S$ from the file output in Algorithm~\ref{alg:solve-jotto}.  This algorithm works by randomly selecting an integer $t$ from 1 to $t^*,$ then playing the guesser's greedy best response to $s_{h,t}$ --- the hider's strategy at iteration $t$.  We can view this algorithm as representing the guesser's strategy as a mixed oracular strategy; in particular, it is the uniform distribution over his greedy best responses in the first $t^*$ iterations of Algorithm~\ref{alg:solve-jotto}. This is noteworthy since it is a rare case of the mixed strategy representation having a computational advantage over the behavioral strategy representation.     

\section{Results}
We ran our algorithm SolveJotto on four different Jotto instances, allowing words to be 2, 3, 4, or 5 letters long.  To speed up the computation, we used the parallel version of the bottleneck subroutine CompNumGuesses (described in Section~\ref{sect:parallel}) with 16 processors.  As our dictionary, we use the Official Tournament and Club Word List~\cite{TWL06}, the official word list for tournament Scrabble in several countries.  As discussed in Section~\ref{sect:Jotto}, we omit words with duplicate letters and words for which there exists an anagram that is also a word.  The dictionary sizes are given in Table~\ref{tab:results}.  We note that our algorithms extend naturally to any number of words and dictionary sizes (and to other variants of Jotto as well).

One metric for evaluating our algorithm is to play the strategies it computes against a benchmark algorithm.  The benchmark algorithm we chose selects his word uniformly at random as the hider, and plays the greedy best response to the uniform strategy as the guesser.  This is the same strategy that we use to initialize our algorithm.

For each number of letters, we computed the payoff of our algorithm SolveJotto against the benchmark (recall that the payoff to the hider is the expected number of guesses needed for the guesser to correctly guess the hider's word). 
The overall payoff is the average of the hider and guesser payoff.  

\begin{table}[!ht]
\begin{center}
\begin{tabular}{c|*{3}{c|}c}
{\bf Number of letters} &2 &3 &4 &5\\ \hline \hline
Dictionary size &51 &421 &1373 &2833\\ \hline
Our hider payoff vs. benchmark &7.652 &7.912 &7.507 &7.221 \\ \hline
Our guesser payoff vs. benchmark &-6.619 &-7.635 &-7.415 &-7.216 \\ \hline
Our overall payoff vs. benchmark &0.517 &0.139 &0.046 &0.003 \\ \hline
Benchmark self-play hider payoff &6.627 &7.601 &7.365 &7.079 \\ \hline
Our algorithm self-play hider payoff &7.438 &7.658 &7.390 &7.162 \\ \hline
Benchmark epsilon &5.373 &3.399 &1.635 &1.921 \\ \hline
Our final epsilon &0.038 &0.334 &0.336 &0.335 \\ \hline
Number of iterations &22212 &10694 &3568 &3906\\ \hline
Avg time per iteration (minutes) &$3.635 \times 10^{-4}$ &0.028 &1.160 &12.576 \\ \hline
\end{tabular}
\caption{Summary of our experimental results.}
\label{tab:results}
\end{center}
\vspace{-0.35in}
\end{table}

Several observations from Table~\ref{tab:results} are noteworthy.  First, our algorithm beats the benchmark for all dictionary sizes.  In the two-letter game, our expected payoff against the benchmark is 0.517; our strategy requires over a full guess less than the benchmark in expectation.  Our profit against the benchmark decreases as more letters are used.

In addition to head-to-head performance against the benchmark, we also compared the algorithms in terms of worst-case performance.  Recall that $\epsilon$ denotes the maximum payoff improvement one player could gain by deviating to a best response (full best response for the hider and greedy best response for the guesser).  Note that in all cases, our $\epsilon$ is significantly lower than that of the benchmark.  For example, in the two-letter game the benchmark obtains an $\epsilon$ of 5.373, while our algorithm obtains one of 0.038. 

Interestingly, we also observe that the self-play payoff of our algorithm, which is an estimate of the value of the game, does not increase monotonically with the number of letters.  That is, increasing the number of letters in the game does not necessarily make it more difficult for the guesser to guess the hider's word.   

\section{Conclusion}
We presented a new approach for computing approximate-equilibrium strategies in Jotto.  Our algorithm produces strategies that significantly outperform a benchmark algorithm with respect to both head-to-head performance and worst-case exploitability.  The algorithm extends fictitious play to a novel strategy representation called oracular form.  We expect our algorithm and the oracular form representation to apply naturally to many other interesting games as well; in particular, games where the strategy space is very large for one player, but relatively small for the other player.

\bibliographystyle{splncs03}
\bibliography{C://Users/Sam/Documents/Research/refs/dairefs}

\end{document}